\documentclass{article}
\usepackage{spconf, amsmath, graphicx}
\usepackage{caption, booktabs} 
\title{PitchNet: Unsupervised Singing Voice Conversion \\with Pitch Adversarial Network}

\name{Chengqi Deng\textsuperscript{1}
    \sthanks{Work performed while interning at Tencent AI Lab},
    Chengzhu Yu\textsuperscript{2}, 
    Heng Lu\textsuperscript{2}, 
    Chao Weng\textsuperscript{2}, 
    Dong Yu\textsuperscript{2}}
\address{\textsuperscript{1}Zhejiang University,             
    \textsuperscript{2}Tencent AI Lab}

\begin{document}
%
\maketitle
\begin{abstract}
Singing voice conversion is to convert a singer's voice to another one's voice without changing singing content. Recent work shows that unsupervised singing voice conversion can be achieved with an autoencoder-based approach \cite{nachmani2019unsupervised}. However, the converted singing voice can be easily out of key, showing that the existing approach cannot model the pitch information precisely.
In this paper, we propose to advance the existing unsupervised singing voice conversion method proposed in \cite{nachmani2019unsupervised} to achieve more accurate pitch translation and flexible pitch manipulation. Specifically, the proposed PitchNet added an adversarially trained pitch regression network to enforce the encoder network to learn pitch invariant phoneme representation, and a separate module to feed pitch extracted from the source audio to the decoder network. 
%
Our evaluation shows that the proposed method can greatly improve the quality of the converted singing voice (2.92 vs 3.75 in MOS). We also demonstrate that the pitch of converted singing can be easily controlled during generation by changing the levels of the extracted pitch before passing it to the decoder network.
\end{abstract}
\section{Introduction}
\label{sec:intro}

Singing is an important way of human expression and the techniques of singing synthesis have broad applications in different prospects including virtual human, movie dubbing and so on. Traditional singing synthesis systems are based on concatenative \cite{bonada2016expressive} or HMM \cite{saino2006hmm} based approaches. With the success of deep learning in Text-to-Speech, some neural singing synthesis methods have also been proposed recently. For example, \cite{blaauw2017neural} introduces a singing synthesis method using an architecture similar to WaveNet \cite{oord2016wavenet}. It adopts lyrics and notes as input and generates vocoder features autoregressively for final singing voice synthesis. 

Singing voice conversion is another way of singing synthesis which extracts musical expression within existing singing and reproduces them with another singer's voice. It is very similar to speech based voice conversion \cite{wu2013exemplar,nakashika2013voice,liu2015video,sun2016phonetic}, but compared with speech voice conversion, singing voice conversion needs to deal with a wider range of frequency variations as well as a sharper change of volume and pitch within singing voice. 
The performance of singing conversion is highly dependent on the musical expression of the converted singing and the similarity of the converted voice timbre compared to the target singer's voice. 

There are several singing voice conversion methods to convert one's singing voice to another \cite{kobayashi2015statistical, kobayashi2014statistical, villavicencio2010applying}. They generally require parallel data to train the conversion model.
To overcome the limitation of the parallel training data for singing voice conversion, an unsupervised method \cite{nachmani2019unsupervised} has been proposed to utilize non-parallel data. This method employs an autoencoder architecture composed of a WaveNet-like encoder, a WaveNet \cite{oord2016wavenet} autoregressive decoder, and a learnable singer embedding table. Voice waveform is passed into the encoder and the output of the encoder will be concatenated with the embedding vector associated with the singer. The concatenated features will be used to condition the WaveNet decoder to reconstruct the input audio. A confusion loss \cite{ganin2016domain} is also introduced to force the encoder to learn a singer-invariant representation. 
By switching among embeddings of different singers during generation, the singing voice conversion can be achieved. While this approach could generate singing voice perceptually similar to the target singer, the quality of generated singing often suffers due to the difficulty of learning a joint representation of phonetic and pitch representation.

To address the difficulty of learning a join phonetic and pitch representation in \cite{nachmani2019unsupervised}, we propose to use adversarially trained pitch regression network to encourage the encoder network to learn not only singer-invariant but also pitch-invariant representation, at the same time extract the pitch from source audio as an additional input to the decoder. The proposed method can greatly improve the quality of the converted voice and achieve flexible pitch manipulation at the same time. 

In the following sections, we will introduce our proposed method in section \ref{sec:method}. And then section \ref{sec:exp} will show that our method is effective by quantitative and qualitative experiments. Finally, we will conclude in section \ref{sec:conclusion} and acknowledgements are in \ref{sec:acknowledgements}.

\section{Method}
\label{sec:method}

\begin{figure}[t]
	\centering{\includegraphics[width=8.0cm]{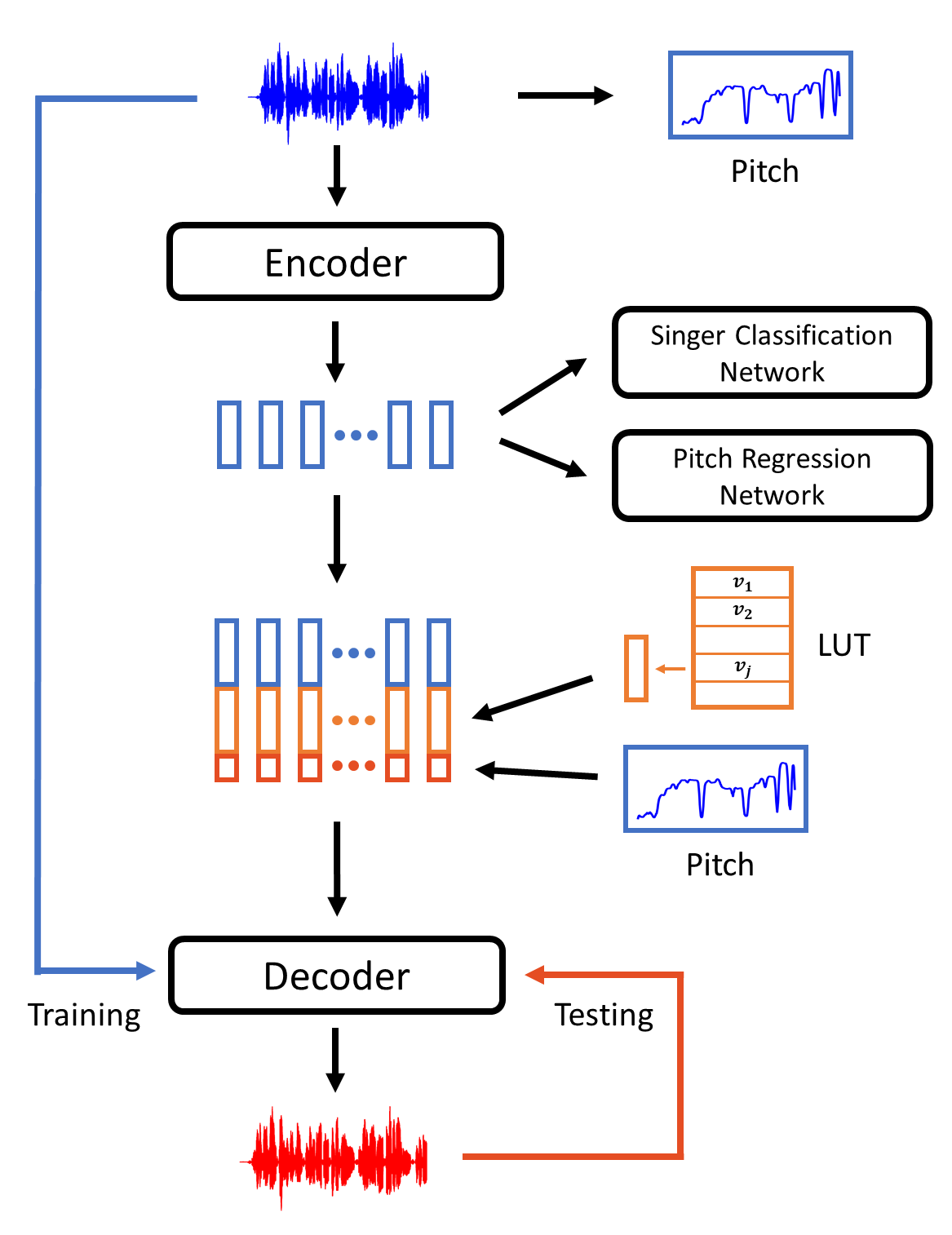}}
	\caption{The overall architecture of PitchNet. PitchNet consists of five parts, an encoder, a decoder, a Look Up Table (LUT) of singer embedding vectors, a singer classification network and a pitch regression network. The audio waveform is directly fed into the encoder. The output of the encoder, the singer embedding vector retrieved from LUT and the input pitch are concatenated together to condition on the WaveNet decoder to output audio waveform.}
	\vspace{-4.5mm}
	\label{fig:arch}
\end{figure}

\begin{figure}[t]
	\centering{\includegraphics[width=8.0cm]{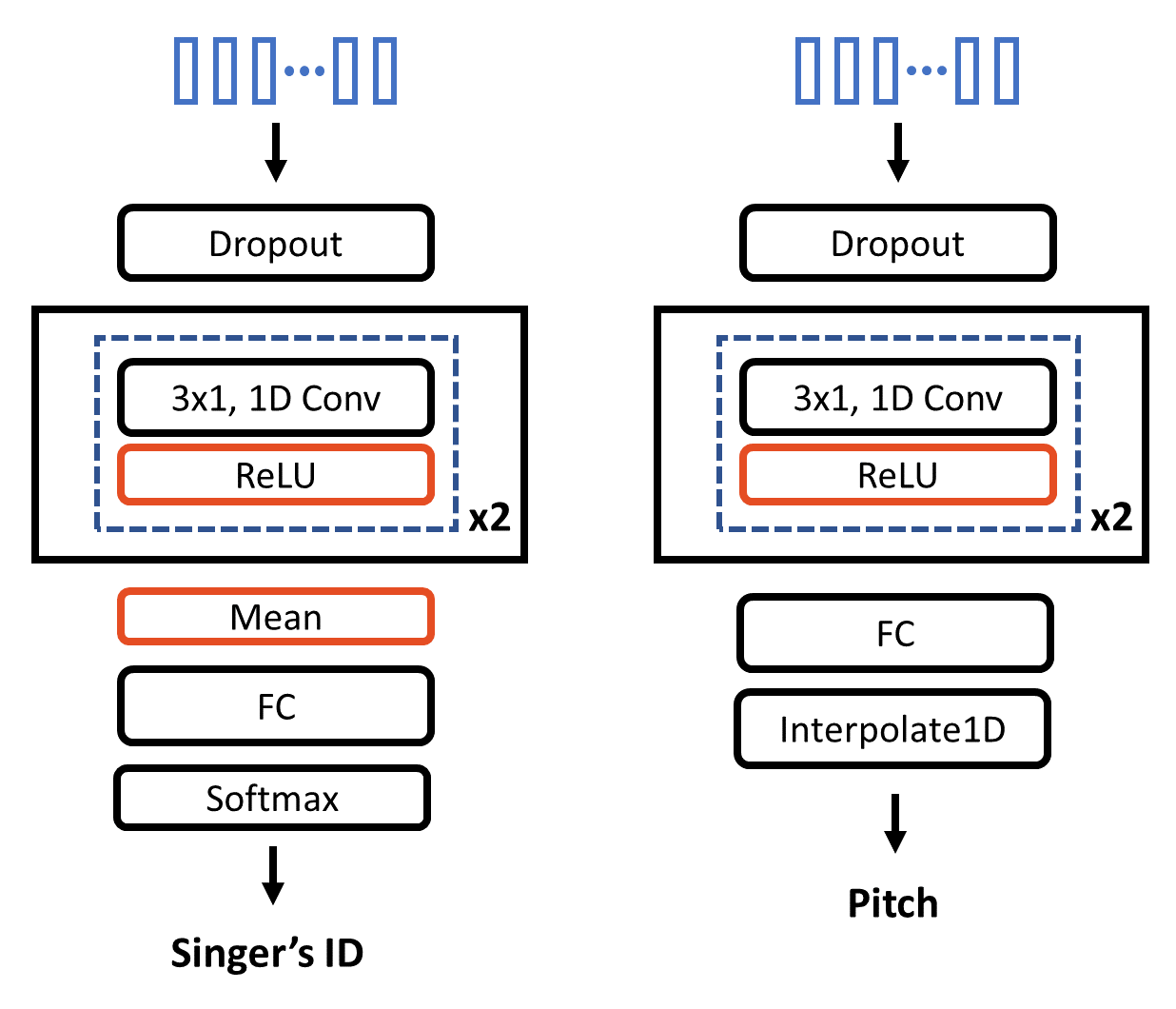}}
	\caption{The architecture of the singer classification network and pitch regression network. Left: Singer classification network; Right: Pitch regression network.}
	\vspace{-4.5mm}
	\label{fig:subarch}
\end{figure}

Our method follows the autoencoder architecture in \cite{nachmani2019unsupervised} except that there is an additional pitch regression network to separate pitch information out of the latent space.
The architecture of PitchNet is illustrated in Fig.~\ref{fig:arch}. It consists of five parts, an encoder, a decoder, a Look Up Table (LUT) of speaker embedding vectors, a singer classification network, and a pitch regression network. 

First, the input waveform is passed through the encoder to extract high-level semantic features. 
An average pooling of stride 800 is then applied to the features, forming a bottleneck to limit the information passing through the encoder. 
After that, a singer id is used to retrieve the target singer's embedding vector from LUT, and concatenated with the output of the encoder at each time step to be a sequence of condition vectors. The pitch of the input audio, extracted separately from the network, is fed into the decoder after a linear interpolation as a compensation signal together with the condition vector. Finally, the decoder is conditioned on the condition vector and pitch to generate audio samples. Since the decoder is an autoregressive model, the output will be fed back to the decoder at the next time step. The model is trained on a softmax-based loss to minimize the reconstruction error with teacher-forcing. 

In order to project the output features of the encoder into a singer and pitch invariant latent space, a singer classification network and a pitch regression network are employed to force the encoder not to encode singer and pitch information. The singer classification loss and pitch regression loss are added adversarially to the reconstruction loss to train the entire model end to end.

\subsection{Training Loss}

To formally describe the model, let $E$ be the encoder network, $D$ be the decoder network, $C_s$ be the singer classification network and $C_p$ be the pitch regression network. Let $v_j$ denote the embedding vector of singer $j$, $s^j$ denote an input audio of singer $j$ and $p(s^j)$ denote the extracted pitch of $s^j$. Now given an input audio sequence $s^j$ and a target singer $k$ where $j,k = 1,2,...,N$ and $N$ is the number of singers, the output of the model is
\begin{equation}
F(s^j, k) = D(
\left[\begin{array}{ccc}
	  	E(s^j)   \\
		p(s^j)   \\
		v_k
	 \end{array} 
\right ]
)
\end{equation}
Note that $D$ is an autoregressive model which would feed the output back to itself.
The reconstruction loss is
\begin{equation}
\mathcal{L}_{recon} = \sum_{j}\sum_{s^j} \mathcal{L}_{ce}(D(
	\left[
	\begin{array}{ccc}
		E(s^j)  \\
		p(s^j)  \\
		v_j
	\end{array} 
	\right ]
), s^j)
\end{equation}
where $\mathcal{L}_{ce}(o, y)$ is the cross entropy loss applied to each element of $o$ and $y$. However, only reconstruction loss is not enough to train the model to learn to convert singing voice between different singers because it just forces the model to reconstruct the input voice.
Therefore, a singer classification loss(also named domain confusion loss \cite{nachmani2019unsupervised}) is applied to make the encoder to learn a singer invariant representation
\begin{equation}
\mathcal{L}_{s} = \sum_{j} \sum_{s^j} \mathcal{L}_{ce}(C_s(E(s^j), j)
\end{equation}
Furthermore, a pitch regression loss is introduced to force the encoder to learn a pitch-independent representation and make the whole model obtain the pitch information from $p(s^j)$ rather than directly from the input audio
\begin{equation}
\mathcal{L}_{p} = \sum_{j} \sum_{s^j} \mathcal{L}_{mse}(C_p(E(s^j), p(s^j))
\end{equation}
where $\mathcal{L}_{mse}(a, b)$ is the mean square error function $\frac{1}{m}||a-b||_2^2$ and m is the number of elements in $a$. The overall loss we minimize to train the model is
\begin{equation}
\label{eq:5}
\mathcal{L}_{total} = \mathcal{L}_{recon} - \lambda \mathcal{L}_{s} - \mu \mathcal{L}_{p}
\end{equation}
where $\lambda$ and $\mu$ are two weight factors.
Furthermore, the adversarial loss used to train the singer classifier and pitch regression network is
\begin{equation} 
\label{eq:6}
\mathcal{L}_{ad} = \lambda \mathcal{L}_{s} + \mu \mathcal{L}_{p}
\end{equation}
In the training process, we minimize $\mathcal{L}_{ad}$ and $\mathcal{L}_{total}$ alternately, that is
\begin{enumerate}
	\item Optimize $C_s$ and $C_p$ one step using $\mathcal{L}_{ad}$ as the objective function.
	\item Optimize the whole model one step using $\mathcal{L}_{total}$ as the objective function.
	\item Go back to step 1.
\end{enumerate}
Furthermore, backtranslation and mixup techniques \cite{nachmani2019unsupervised} are also used to improve the quality of the converted singing voice.

\subsection{The architecture of the Sub-Networks}
The encoder and decoder networks follow the design in \cite{nachmani2019unsupervised} which is already shown to be effective for singing voice conversion. The encoder is a fully convolutional network with three blocks of ten residual-layers which consists of a ReLU activation, a dilated convolution, a ReLU activation, a 1x1 convolution, and a residual summation in order. After three residual blocks, a 1x1 convolution and an average pooling with a kernel size of 800 are applied to get the final output. The decoder is a WaveNet \cite{oord2016wavenet} vocoder which consists of four blocks of ten residual layers. The linear interpolation and nearest-neighbor interpolation are applied to the input pitch and encoder output respectively, upsampling them to be of the same sample rate as the input audio waveform.

As shown by Fig.~\ref{fig:subarch}, the singer classification network and pitch regression network have the same architecture of a stack of two convolutional neural networks with a kernel size of 3 and channels of 100. Except that the pitch regression network does not average the output of the two convolution networks along the time dimension before passing it into the final fully connected network. A dropout layer is also employed at the beginning of the network to make the training process more stable.

\section{Experiments}
\label{sec:exp}

\begin{table}[t]
\caption{Automatic quality scores}
	\begin{center}
		\label{tab:ncc}
		\begin{tabular}{lcc}
			\toprule[1.5pt]
			\textbf{Method} & \textbf{NCC score}\\
			\midrule
			USVC (Our, reconstruction) & 0.838\\
			USVC (Our, conversion) & 0.821\\
			PitchNet (reconstruction) & 0.882\\
			PitchNet (conversion) & 0.855\\
			\bottomrule[1.2pt]
		\end{tabular}
	\end{center}
\end{table}

\begin{table}[t]
\caption{MOS scores}
	\begin{center}
		\label{tab:mos}
		\begin{tabular}{lcc}
			\toprule[1.5pt]
			\textbf{Method} & \textbf{Naturalness} & \textbf{Similarity}\\
			\midrule
			USVC (Original) & 3.06 & 3.34 \\
			USVC (Our) & 2.92 & 3.14 \\
			PitchNet & \textbf{3.75} & \textbf{3.64} \\
			\bottomrule[1.2pt]
		\end{tabular}
	\end{center}
\end{table}

\begin{figure}[t]
	\centering{\includegraphics[width=8.5cm]{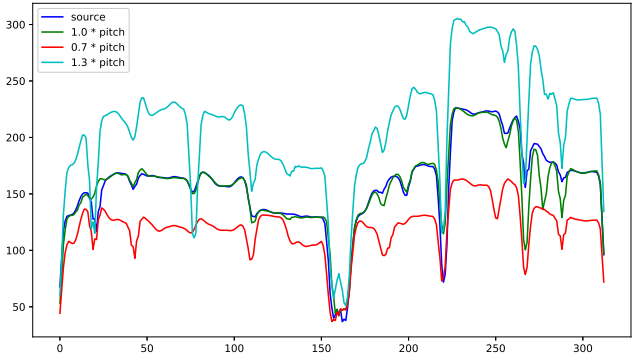}}
	\caption{The pitch of the source audio and converted audio with different pitch as input}
	\label{fig:pitch-change}
\end{figure}

Here we compare the audio quality between our method and \cite{nachmani2019unsupervised}'s method (Below we call USVC) and show that the input pitch can affect the output singing voice by qualitative analysis.
Since the authors of \cite{nachmani2019unsupervised} do not release their source code and only provide part of the converted results at their website, we implemented USVC by ourselves, denoted by USVC(our) below, to give a more comprehensive comparison. 
Audio samples are available at our website \footnote{https://tencent-ailab.github.io/pitch-net/}.

\subsection{Dataset and Preprocessing}
NUS-48E \cite{duan2013nus} dataset, sung by 6 male singers and 6 female singers, was used to train the models. It contains 48 songs each with a length of several minutes. Every singer provided 4 songs.
The male part of the dataset was selected to train the models. During testing, We converted each one's singing voice to the other five singer's voice.
Before training, We converted the songs to monophonic audio of 16kHz sample rate and PCM-16 bit format. Besides, 8-bit mu-law encoding was employed to reduce the input space to speed up the training process, although it will degrade the audio quality. Kaldi toolkit \cite{povey2011kaldi} was used to extract pitch from the songs with hop length of 100 which means that we could get 1600 pitch samples in an audio segment of one second. Before feeding them into the model, we normalized the value of pitch between 0 and 1.

\subsection{Training}
We implemented USVC and PitchNet using PyTorch \cite{paszke2017automatic} framework.
Both models were trained on two Tesla P40 GPUs for four days. Adam optimizer \cite{kingma2014adam} was used with a learning rate of $10^{-3}$ and a decay factor of 0.98 every 1000 steps. 
The models were trained up to 30k steps with a batch size of 4. $\lambda$ and $\mu$ in the training loss \eqref{eq:5}\eqref{eq:6} were set to $0.01$ and $0.1$ respectively. The dropout probability in the singer classification network and pitch regression network were both $0.2$. 

During the training process, backtranslation and mixup \cite{nachmani2019unsupervised} were employed to improve the conversion. New training samples were generated by mixing embedding vectors of two different singers A and B with a uniform random weight factor. Then these samples were fed to the model to reconstruct A's voice with the embedding vector of A. The reconstructed voice and original voice were used to calculate the reconstruction loss. After training for 200k steps without backtranslation and mixup, we generated 96 new audio segments every 2k steps and used them to train for 24 steps without the adversarial loss \eqref{eq:6}. 

Besides, audio time reversal and phase inversion \cite{nachmani2019unsupervised} were also employed to augment the training data by 4 times. 

\subsection{Evaluation}
To compare the conversions between USVC and PitchNet, we employed an automatic evaluation score and a human evaluation score. 

The automatic score roughly followed the design in \cite{mor2018universal}. The pitch tracker of librosa package \cite{mcfee2015librosa} was employed to extract pitch information of the input and output audio. Then the output pitch was compared to the input pitch using the normalized cross correlation (NCC) which would give a score between 0 and 1. The higher the score is, the better the output pitch matches the input pitch. We conducted the evaluation on USVC (our) and PitchNet. The evaluated automatic scores on conversion and reconstruction tasks are shown in Tab.~\ref{tab:ncc}. Our method performed better both on conversion and reconstruction. The scores of reconstruction are higher than conversion since both models were trained using a reconstruction loss. However, the score of our method on conversion is even higher than the score of USVC (Our) on reconstruction.

Mean Opinion Score (MOS) was used as a subjective metric to evaluate the quality of the converted audio. Two questions were asked: (1) what is the quality of the audio? (naturalness) (2) How well does the converted version match the original? (similarity) A score of 1-5 would be given to answer the questions. The evaluation was conducted on USVC (Our) and PitchNet. Besides, the converted samples provided by \cite{nachmani2019unsupervised} was also included to give a more convincing evaluation. As shown by Tab.~\ref{tab:mos}, the naturalness and similarity of our method are both higher than the other two ones. Our implementation of USVC performed slightly lower than the original author's because we cannot fully reproduce the results of them.

Next we qualitatively analyze the influence of input pitch in our method. We used different pitch as input to observe how the output pitch would change along with the input pitch. The input pitch was multiplied by 0.7, 1.0 and 1.3 respectively. And the output pitch was also extracted by the pitch tracker of the librosa package. Fig.~\ref{fig:pitch-change} plots the pitch of input audio and output audio with different pitch as input while keeping the target singer the same. As shown by Fig.~\ref{fig:pitch-change}, the output pitch changes significantly along with the input pitch. The examples are also presented at our website.

\section{Conclusion}
\label{sec:conclusion}

In this paper, a novel unsupervised singing voice conversion method named PitchNet is proposed. A pitch regression network is employed to render an adversarial loss separating pitch related information from the latent space in autoencoder. After the WaveNet-like encoder, a singer and pitch invariant representation is generated and then fed into the WaveNet decoder conditioning on the singer embedding and the extracted pitch to reconstruct the target singing voice. Our method outperforms the existing unsupervised singing voice conversion method and achieves flexible pitch manipulation. 

\section{Acknowledgements}
\label{sec:acknowledgements}
The authors would like to thanks Kun Xu and other members in the Tencent AI Lab team for discussions and suggestions.


\bibliographystyle{IEEEbib}
\bibliography{refs}

\end{document}